\documentclass[3p,times,procedia]{elsarticle}
\usepackage{nupha_ecrc}

\volume{00}

\firstpage{1}

\journalname{Nuclear Physics A}

\runauth{}

\jid{nupha}

\jnltitlelogo{Nuclear Physics A}

\usepackage{amssymb}
\usepackage[figuresright]{rotating}

\begin{document}

\begin{frontmatter}



%% Instructions from Editor: Please use the following \dochead only in the preprint version (e-print arXiv etc.); 
%% use empty \dochead{} when submitting to Nuclear Physics A!
\dochead{XXVIIIth International Conference on Ultrarelativistic Nucleus-Nucleus Collisions\\ (Quark Matter 2019)}
%% e.g. \dochead{17th International Conference on Dynamical Processes in Excited States of Solids}

\title{Energy dependence of longitudinal flow decorrelation from STAR}


\author{Maowu Nie for the STAR Collaboration}

\address{Institute of Frontier and Interdisciplinary Science, Shandong University, Qingdao, Shandong, 266237, China}
\address{Key Laboratory of Particle Physics and Particle Irradiation, Ministry of Education, Shandong University, Qingdao, Shandong, 266237, China}
\address{Department of Chemistry, Stony Brook University, Stony Brook, NY 11794, USA}

\begin{abstract}

Measurements of longitudinal flow decorrelations for charged particles are presented in the pseudorapidity range $|\eta| < 1$ using a reference detector at 2.1 $< |\eta_{\mathrm{ref}}| <$ 5.1 in Au+Au collisions at $\sqrt{s_{NN}}$ = 27 GeV by STAR.  The flow decorrelation for $v_2$ shows a strong centrality dependence, while a weak centrality dependence for $v_3$. Results are compared with the results in Au+Au collisions at 200 GeV as a function of $\eta$ scaled by beam-rapidity, i.e. $\eta/y_{beam}$. No energy dependence is observed for $v_2$  decorrelation, but clear energy dependence for $v_3$ decorrelation. These results provide new insights into the longitudinal structure of the initial-state geometry in heavy-ion collisions.

\end{abstract}

\begin{keyword}
flow decorrelation, longitudinal dynamics, STAR
\end{keyword}

\end{frontmatter}

%%\linenumbers

\section{Introduction}
\label{char:intro}
Initial-state fluctuations in the transverse plane in the heavy-ion collisions play an important role for the final-state dynamics of multiparton interactions in Quark-Gluon Plasma. Recently, it is realized that the longitudinal fluctuations are also important for the longitudinal dynamics of the medium evolution produced in heavy-ion collisions~\cite{Bozek:2010vz,Jia:2014ysa}. The rapidity decorrelation of flow harmonics, $\mathbf V_n$, probes the non-boost-invariant nature of the initial-collision geometry and final-state collective dynamics~\cite{Jia:2017kdq}. In this proceeding, we present the new measurements of flow decorrelation in Au+Au at $\sqrt{s_{NN}}$ = 27 GeV with the STAR detector. Comparisons with results from top RHIC energy and the LHC energies are discussed.

\section{Analysis method}
\label{char:method}

The azimuthal anisotropy of the particle production in an event is described by harmonic flow vector, $\mathbf V_n = v_ne^{in\Psi_{n}}$, where $v_n$ and  $\Psi_n$ are the magnitude and phase (event plane) of $n^{th}$-order flow harmonics, respectively. Experimentally, $\mathbf V_n$ is estimated from the observed per-particle flow vector, ${\mathbf q_{n}}  \equiv  \sum \omega_{i} e^{in\phi_{i}}/{\sum \omega_{i}}$, where the sum runs over all charged particles in the phase-space sample and $\omega_{i}$ is the weight assigned to the $i^{th}$ particle. The $\mathbf q_{n}$ deviates from $\mathbf V_{n}$  due to non-flow contribution and statistical fluctuations. The non-flow contribution can be effectively suppressed by requiring a large pseudorapidity gap, and the statistical fluctuation drops after event average. The correlation between $\mathbf V_n$ from two pseudorapidity intervals can then be estimated with the observed flow vector $\mathbf q_{n}$:

\begin{eqnarray}
	\label{eq: 2PC}
	\langle {\mathbf{q}_n ( {\eta})  \mathbf{q}^*_n(\eta_{\mathrm{ref}})} \rangle = \langle {\mathbf{V}_n ( {\eta})  \mathbf{V}^*_n(\eta_{\mathrm{ref}})} \rangle 
\end{eqnarray}

The flow decorrelations are quantified with the factorization ratio, $r_n$, which is proposed by the CMS experiments\cite{Khachatryan:2015oea}. Since $\langle {\mathbf{q}_n ( {-\eta})  \mathbf{q}^*_n(\eta_{\mathrm{ref}})} \rangle   =\langle {\mathbf{q}_n ( {\eta})  \mathbf{q}^*_n(-\eta_{\mathrm{ref}})} \rangle$, for a symmetric system, a symmetrization procedure has been applied to further cancel the possible differences between $\eta$ and $-\eta$ in the tracking efficiency or detector acceptance. The observable, as shown in Eq. \ref{eq: ratio}, is sensitive to the event-by-event fluctuations of the initial conditions in the longitudinal direction. If flow harmonics from two-particle correlations factorize into single-particle flow harmonics, then the value of $r_n$ is expected to be equal to unity. Therefore, $r_n \neq 1$ would indicate the effects of longitudinal flow decorrelations.

\begin{eqnarray}
	\label{eq: ratio}
	r_{n} (\eta) =  \frac{ \langle {\mathbf{q}_n ( {-\eta})  \mathbf{q}^*_n(\eta_{\mathrm{ref}})} +{\mathbf{q}_n ( {\eta})  \mathbf{q}^*_n(-\eta_{\mathrm{ref}})} \rangle } { \langle {\mathbf{q}_n ( {\eta})  \mathbf{q}^*_n(\eta_{\mathrm{ref}})} +{\mathbf{q}_n ( {-\eta})  \mathbf{q}^*_n(-\eta_{\mathrm{ref}})}  \rangle}\\ \nonumber
\end{eqnarray}

In this analysis, the measurements are performed using charged particles with $0.4 < p_{\mathrm{T}} < 4$  GeV/c from the Time Projection Chamber (TPC, $|\eta|<1$), and the reference flow vector is calculated from the Event Plane Detector (EPD, $2.1<|\eta_{\mathrm{ref}}|<5.1$) for the $\sqrt{s_{NN}}$ = 27 GeV Au+Au collisions. The systematic uncertainties sources are estimated using positive tracks, negative tracks and tight track selections.

\section{Results and discussion}
\label{char:results}

\begin{figure}[!htbp]
	\centering
	\includegraphics[width=0.4\textwidth]{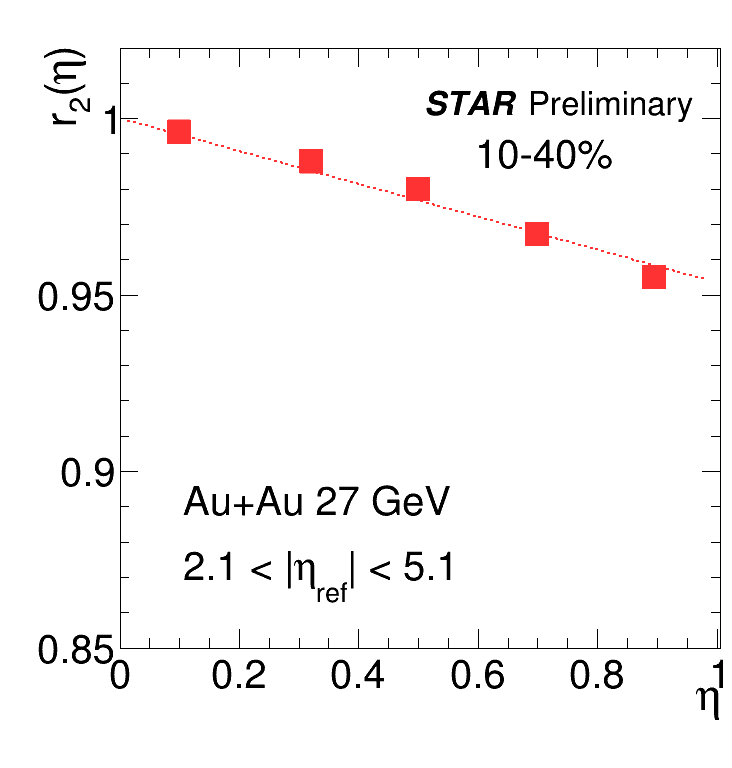}
	\includegraphics[width=0.4\textwidth]{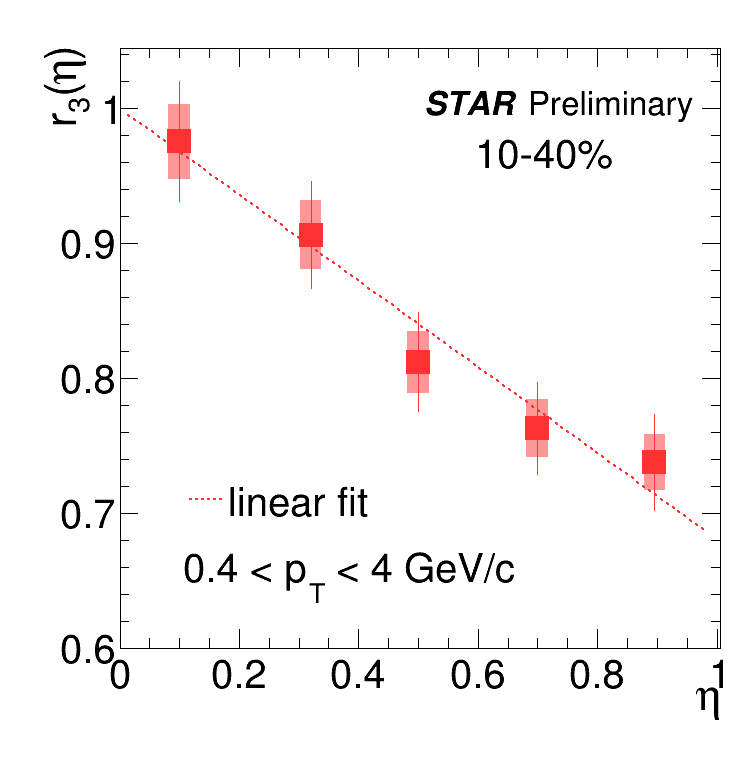}
	\caption{The factorization ratios, $r_2$ (left panel) and $r_3$ (right panel), as a function of $\eta$, averaged over $0.4 < p_{\mathrm{T}} < 4 $ GeV/c Au+Au collisions, The error bars and solid boxes are statistical and systematic uncertainties, respectively.}
	\label{fig:r2andr3}
\end{figure}

Figure \ref{fig:r2andr3} shows the factorization ratios $r_2$ and $r_3$ as a function of $\eta$, averaged over $0.4 < p_{\mathrm{T}} < 4$ GeV/c for 10-40\% Au+Au collisions. Both $r_2$ and $r_3$ decrease linearly with increasing $\eta$. The decreasing trend can be described by a linear fit. The similar behavior is also observed for other centralities.

\begin{figure}[!htbp]
	\centering
	\includegraphics[width=0.8\textwidth]{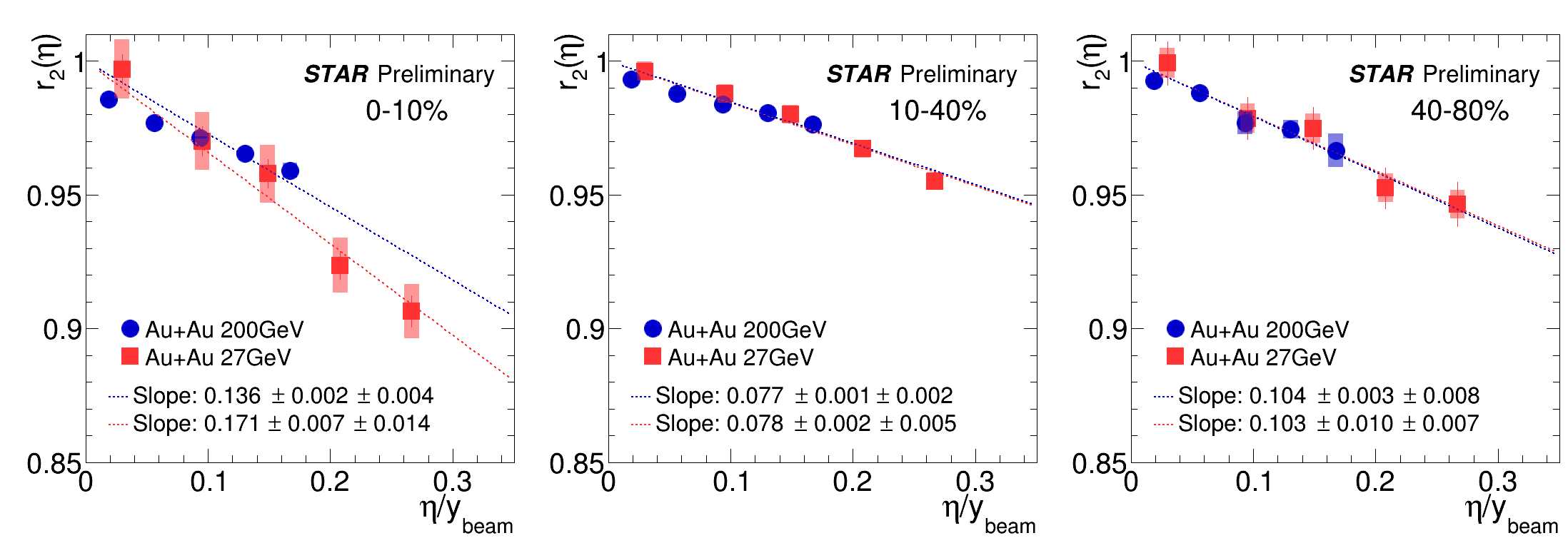}
	\includegraphics[width=0.8\textwidth]{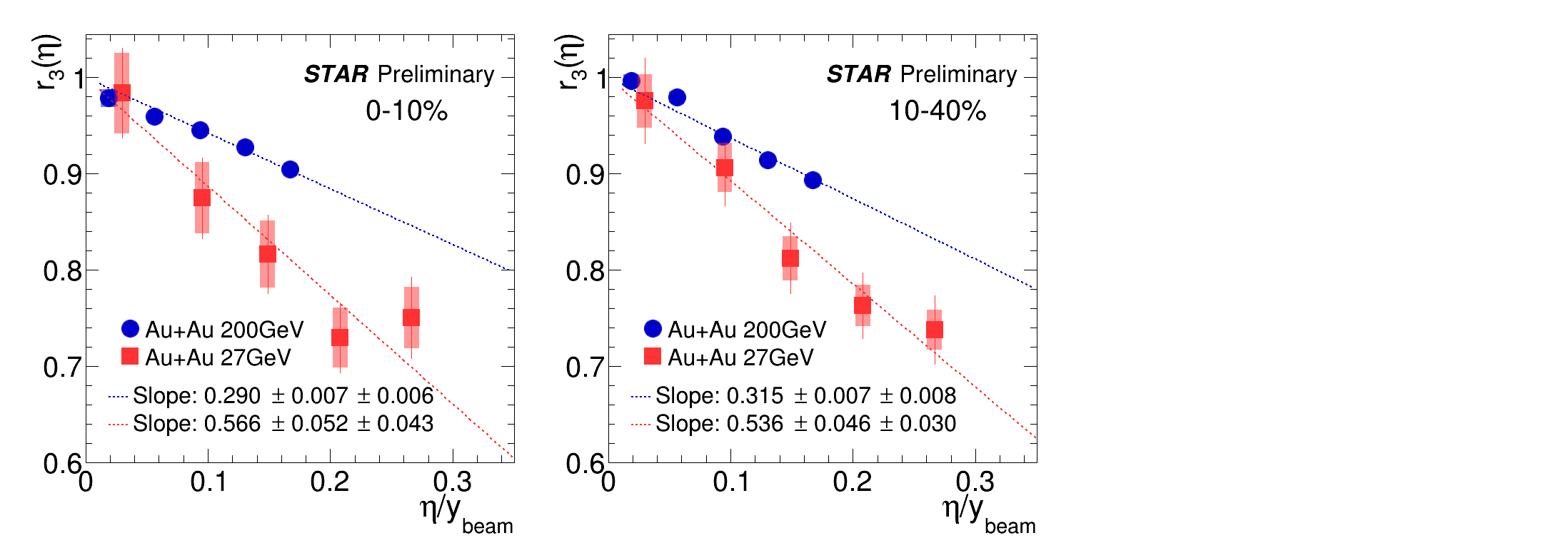}
	\caption{The $r_2(\eta)$ (top panel) and $r_3(\eta)$ (bottom panel) as a function of $\eta/y_{\mathrm{beam}}$ in three centrality bins, the comparison between Au+Au 27~GeV (red marker) and Au+Au 200~GeV (blue marker) are also shown.}
	\label{fig:compare200}
\end{figure}

The first measurement of longitudinal decorrelations of harmonic flow $v_2$ and $v_3$ in Au+Au collisions at $\sqrt{s_{NN}}$ = 200 GeV has been reported by STAR in Ref. ~\cite{Nie:2019bgd}. A direct comparison of decorrelation results between 27 and 200 GeV can help us to better understand the energy dependence of the longitudinal dynamics.  To account for the beam-rapidity dependence, a rapidity normalization procedure is applied for the comparison. Figure \ref{fig:compare200} shows $r_n$ as a function of normalized pseudorapidity $\eta/y_{\mathrm{beam}}$, where $y_{\mathrm{beam}}$ = 5.36 for 200 GeV and 3.36 for 27 GeV. The factorization ratio, $r_2$ (top panel), is plotted in various centrality intervals. The filled squares and circles are the results for 27 and 200 GeV, respectively. The $v_2$ decorrelation at 27 GeV is slightly stronger in $0-10\%$ than 200GeV, but are nearly the same between the two energies for other centrality ranges. The results suggest no clear energy dependence of $r_2$ after beam-rapidity normalization. For the factorization ratio, $r_3$ (bottom panel), on the other hand,  shows weak centrality dependence. The $r_3$ shows a clear energy dependence and a stronger decorrelation for 27 GeV.

% To study the centrality dependence, the $r_n$ is parameterized with a linear function, $r_n = 1 - 2F_n \eta$. Figure \ref{fig:compare200} shows the centrality dependence of $F_n$ in terms of $N_\mathrm{part}$, which quantifies the strength of the decorrelation effect. $F_2$ shows a clearly centrality dependence, where the decorrelation effect is the weakest at midcentral collisions. $F_3$ shows weak centrality dependence. 
%The results at $\sqrt{s_{NN}}= $200~GeV and $\sqrt{s_{NN}}= $5.02~TeV show the strongest and the weakest decorrelation effects, respectively.

\begin{figure}[!htbp]
	\centering
	\includegraphics[width=0.4\textwidth]{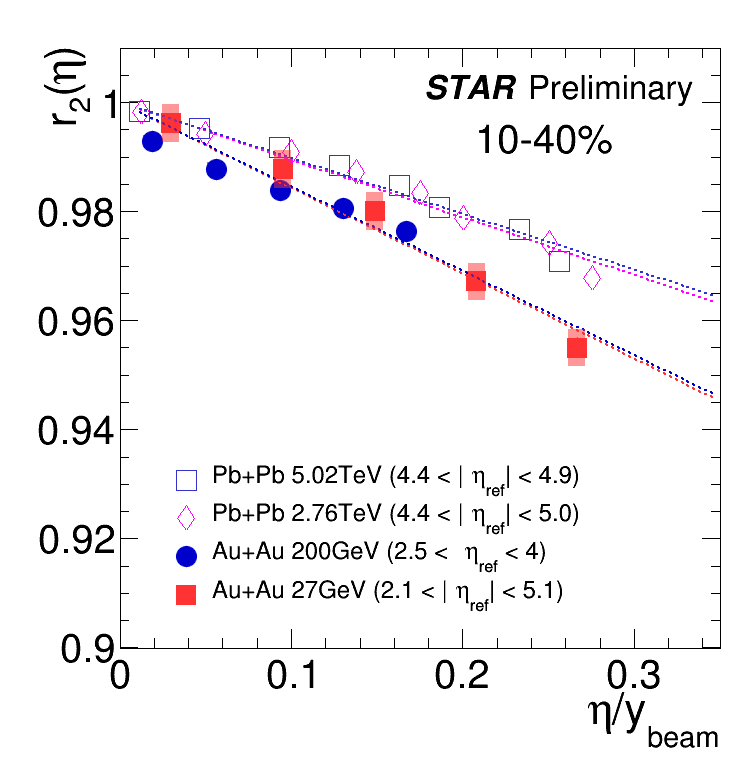}
	\includegraphics[width=0.4\textwidth]{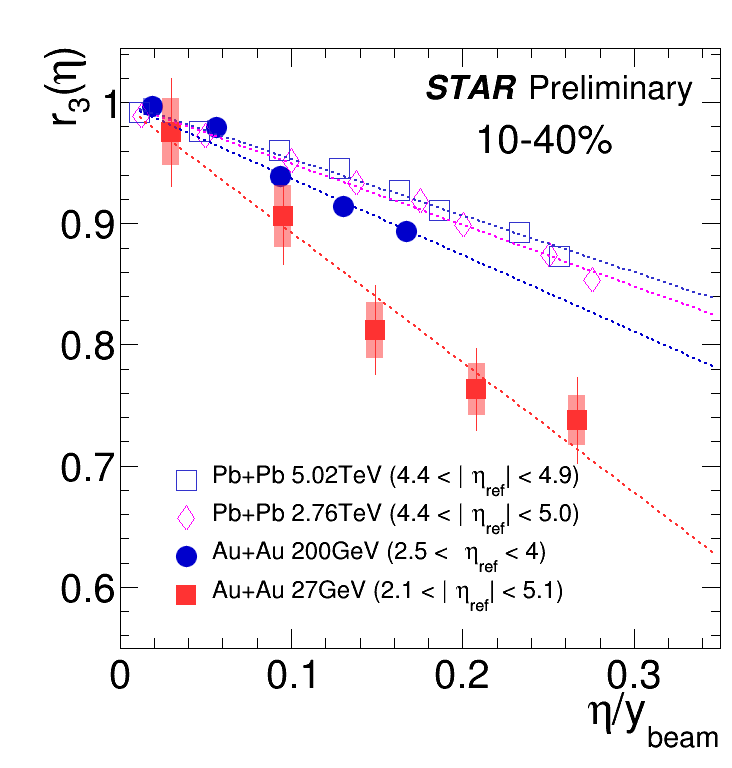}
	\caption{The $r_2(\eta)$ (left panel) and $r_3(\eta)$ (right panel) as a function of $\eta/y_{\mathrm{beam}}$ in 10-40\%, and compared with four collision energies.}
	\label{fig:compareLHC}
\end{figure}

The previous comparison between results at the top RHIC energies and at the LHC energies~\cite{Khachatryan:2015oea,Aaboud:2017tql} indicate stronger decorrelation effect at lower energies. To investigate the energy dependence of flow decorrleation, we further compare three other energies with the new measurements, as shows in Figure \ref{fig:compareLHC}. For $r_2$, we find neither RHIC nor LHC energies had clear energy dependence after beam-rapidity normalization. The results show the decorrelation is stronger at RHIC energies than at the LHC energy. This results is still not understood and need further studies in both experimental and theoretical studies. The high statistics Au+Au collisions at $\sqrt{s_{NN}}$ = 54.4 GeV will be used to test the energy dependence, and the future RHIC Beam Energy Scan II(BES-II) data are also crucial for this study. On the other hand, the $r_3$ shows clear energy depdence, lower energy has stronger decorrelation effect. The observed energy dependence of $r_2$ and $r_3$ remain to be a puzzle from current understanding, futher studies are still needed.

%\begin{figure}[!htbp]
%	\centering
%	\includegraphics[width=1.0\textwidth]{./3can_r3.png}
%	\caption{Same as Figure \ref{fig:r2}, but for $r_3(\eta)$.}
%	\label{fig:r3}
%\end{figure}

\section{Conclusions}
\label{char:conclusion}

Measurements of longitudinal flow correlations for charged particles are presented in the pseudorapidity range $|\eta| < 1$ using a reference detector at $2.1 < \eta_{\mathrm{ref}} < 5.1$ in Au+Au collisions at $\sqrt{s_{NN}}$ = 27 GeV with the STAR detector at RHIC. The strength of the decorrelation is nearly independent of centrality for $r_3$. However, for $r_2$ the effect has a strong centrality dependence. The results are compared with those from LHC and top RHIC energies. After beam-rapidity normalization, the $r_2$ shows no clear energy dependence for both RHIC or LHC energies, while the $r_3$ show clear hierarchy and the decorrelation effect is stronger at lower energy. 

\section{Acknowledgements}
\label{char:acknowledgements}

This work is supported by the China Postdoctoral Science Foundation 2019M662319, NSFC grant number 11890713, PHY-1613294, PHY-1913138, and the Program of Qilu Young Scholars of Shandong University.



%% References


\bibliographystyle{elsarticle-num}
\bibliography{ref}







% \end{thebibliography}

\end{document}